\documentclass[onecolumn]{aa}  
\usepackage{graphicx}
\usepackage{listings}
\usepackage{xspace}
\usepackage{natbib}
\usepackage{url}
\usepackage[xindy, toc, hyperfirst=false, nolist, nostyles, sanitize={name=false,description=false,symbol=false}]{glossaries}
\bibpunct{(}{)}{;}{a}{}{,} 
\usepackage{graphicx}
\usepackage{txfonts}
\begin{document} 

   \title{EinsteinPy: A Community Python Package for General Relativity}

   \subtitle{}

   \author{
Shreyas Bapat\inst{\ref{inst:iitm}}  
  \and
Ritwik Saha\inst{\ref{inst:iitm}}  
  \and
Bhavya Bhatt\inst{\ref{inst:iitm}}  
\and
Shilpi Jain\inst{\ref{inst:iitr}}
\and
Akshita Jain\inst{\ref{inst:iitm}}
\and
Sofía Ortín Vela\inst{\ref{inst:uz}}
\and
Priyanshu Khandelwal\inst{\ref{inst:iitm}}
\and
Jyotirmaya Shivottam\inst{\ref{inst:niser}}
\and
Jialin Ma\inst{\ref{inst:gatech}}
\and
Gim Seng Ng\inst{\ref{inst:trinity}}
\and
Pratyush Kerhalkar\inst{\ref{inst:manipal}}
\and
Hrishikesh Sudam Sarode\inst{\ref{inst:iitm}}
\and
Rishi Sharma\inst{\ref{inst:iitm}}
\and
Manvi Gupta\inst{\ref{inst:iitm}}
\and
Divya Gupta\inst{\ref{inst:iitm}}
\and
Tushar Tyagi\inst{\ref{inst:iitm}}
\and
Tanmay Rustagi\inst{\ref{inst:iitm}}
\and
Varun Singh\inst{\ref{inst:iitm}}
\and
Saurabh Bansal\inst{\ref{inst:iitm}}
\and
Naman Tayal\inst{\ref{inst:iitm}}
\and
Abhijeet Manhas\inst{\ref{inst:iitm}}
\and
Raphael Reyna\inst{\ref{inst:calpoly}}
\and
Gaurav Kumar\inst{\ref{inst:iitm}}
\and
Govind Dixit\inst{\ref{inst:iiitl}}
\and
Ratin Kumar\inst{\ref{inst:nitk}}
\and
Sashank Mishra\inst{\ref{inst:iiita}}
\and
Alpesh Jamgade\inst{\ref{inst:bharath}}
\and
Raahul Singh\inst{\ref{inst:sri}}
\and
Rohit Sanjay\inst{\ref{inst:manipal}}
\and
Khalid Shaikh\inst{\ref{inst:yadav}}
\and
Bhavam Vidyarthi\inst{\ref{inst:manipal}}
\and
Shamanth R Nayak K\inst{\ref{inst:manipal}}
\and
Vineet Gandham\inst{\ref{inst:manipal}}
\and
Nimesh Vashistha\inst{\ref{inst:manipal}}
\and
Arnav Das\inst{\ref{inst:hira}}
\and
Saurabh\inst{\ref{inst:du}}
\and
Shreyas Kalvankar\inst{\ref{inst:wagh}}
\and
Ganesh Tarone\inst{\ref{inst:vish}}
\and
Atul Mangat\inst{\ref{inst:iitr}}
\and
Suyog Garg\inst{\ref{inst:iiitdm}}
\and
Bibek Gautam\inst{\ref{inst:xav}}
\and
Sitara Srinivasan\inst{\ref{inst:nax}}
\and
Aayush Gautam\inst{\ref{inst:tribhu}}
\and
Swaastick Kumar Singh\inst{\ref{inst:amity}}
\and
Suyash Salampuria\inst{\ref{inst:iitr}}
\and
Zac Yauney\inst{\ref{inst:bri}}
\and
Nihar Gupte\inst{\ref{inst:uf}}
\and
Gagan Shenoy\inst{\ref{inst:manipal}}
\and
Micky Yun Chan\inst{\ref{inst:lat}}
}
   \institute{
  Indian Institute of Technology Mandi, Mandi, Himachal Pradesh, India 175005
  \label{inst:iitm}
    \and
  Indian Institute of Technology Roorkee, Roorkee, Uttarakhand, India
  \label{inst:iitr}
    \and
  University of Zaragoza, Calle de Pedro Cerbuna, 12, 50009 Zaragoza, Spain
  \label{inst:uz}
    \and
  Georgia Institute of Technology, USA
  \label{inst:gatech}
  \and
   School of Mathematics and Hamilton Mathematics Institute, Trinity College Dublin, Dublin 2, Ireland  
   \label{inst:trinity} 
    \and
   School of Physical Sciences, National Institute of Science Education and Research, Jatni, Khurdha, Odisha, India
   \label{inst:niser}
   \and
   Department of Electronics and Communication Engineering, Manipal Institute of Technology, India
   \label{inst:elecmani}
   \and
   California State Polytechnic University, Pomona, USA
   \label{inst:calpoly}
   \and
   Department of Information Technology, Indian Institute of Information Technology Lucknow, India
   \label{inst:iiitl}
   \and
   National Institute of Technology Kurukshetra, India
   \label{inst:nitk}
   \and
   Department of Information Technology, Indian Institute of Information Technology Allahabad, India
   \label{inst:iiita} 
   \and
   Department of Mathematics, Bharath Institute of Higher Education and Research, Chennai, India
   \label{inst:bharath}  
   \and
   Indian Institute of Information Technology, Sri City, India
   \label{inst:sri} 
   \and
   Yadavrao Taskgaonkar Institute of Engineering and Technology, Mumbai University, Mumbai, Maharashtra, India
   \label{inst:yadav}
   \and
   Manipal Institute of Technology, Manipal, Karnataka, India
   \label{inst:manipal}
   \and
   Hiranandani Foundation, Powai, Mumbai, Maharashtra, India
   \label{inst:hira}
   \and
   Department of Physics, Dyal Singh College, University of Delhi, New Delhi, India  
   \label{inst:du} 
   \and
   K. K. Wagh Institute of Engineering Education and Research, Nashik, Maharashtra, India  
   \label{inst:wagh}
    \and
   Department of Computer Engineering,Vishwakarma Institute of Technology,Pune, Maharashtra, India  
   \label{inst:vish} 
  \and
  Indian Institute of Information Technology, Design and Manufacturing, Chennai, Tamil Nadu, India 600127
  \label{inst:iiitdm}
  \and
  Naxxatra Sciences and Collaborative Research, Bangalore, Karnataka, India 
  \label{inst:nax}
  \and
  St. Xavier's College, Maitighar, Kathmandu, Nepal
  \label{inst:xav}
  \and
  Amity University Kolkata, West Bengal, India 
  \label{inst:amity}
  \and
  Department of Physics, Birendra Multiple Campus, Tribhuwan University, Kathmandu, Nepal
  \label{inst:tribhu}
  \and
  Brigham Young University, Provo, UT 84602, United States
  \label{inst:bri}
  \and
  Department of Physics, University of Florida, Gainesville, FL, USA
  \label{inst:uf}
  \and
  Latvia University of Life Sciences and Technologies, Lielā iela 2, Jelgava, LV-3001, Latvia
  \label{inst:lat}
}

   \date{Received May 21, 2020; accepted May 21, 2020}

  \abstract
{This paper presents EinsteinPy (version 0.3), a community-developed Python package for gravitational and relativistic astrophysics. Python is a free, easy to use a high-level programming language which has seen a huge expansion in the number of its users and developers in recent years. Specifically, a lot of recent studies show that the use of Python in Astrophysics and general physics has increased exponentially. We aim to provide a very high level of abstraction, an easy to use interface and pleasing user experience. EinsteinPy is developed keeping in mind the state of a theoretical gravitational physicist with little or no background in computer programming and trying to work in the field of numerical relativity or trying to use simulations in their research. Currently, EinsteinPy supports simulation of time-like and null geodesics and calculates trajectories in different background geometries some of which are Schwarzschild, Kerr, and KerrNewmann along with coordinate inter-conversion pipeline. It has a partially developed pipeline for plotting and visualization with dependencies on libraries like Plotly, matplotlib, etc. One of the unique features of EinsteinPy is a sufficiently developed symbolic tensor manipulation utilities which are a great tool in itself for teaching yourself tensor algebra which for many beginner students can be overwhelmingly tricky. EinsteinPy also provides few utility functions for hypersurface embedding of Schwarzschild spacetime which further will be extended to model gravitational lensing simulation.}

   \keywords{
    gravitational physics --
    astrophysics --
    simulations --
    black holes --
    astropy
    }

   \maketitle
%

\section{Introduction}

   The relativistic theory of gravity paved its way to the ground in 1916, 
   when Prof. Albert Einstein published his paper on the general theory of 
   relativity \citeyear{Einstein:1916vd}. This elegant and rigorous framework 
   was a generalized version of gravity-free theory - special relativity, 
   which he published earlier in 1905. Even after almost a hundred years 
   after Einstein wrote down the equations of General Relativity, solutions 
   of the Einstein field equations remain extremely difficult to find beyond 
   those which exhibit significant symmetries.

   We can study the behaviour of solutions under a high degree of symmetry 
   considerations and could even solve analytically for highly unrealistic 
   systems. In case of problems relevant to astrophysical and gravitational 
   physics research, it remains a big daunting question on how to get around 
   the problem of solving these field equations. This question is so profound 
   that it has a separate field of research which goes with the name of numerical  
   relativity and attempts to use computation science and programming to numerically 
   obtain solutions of the equations (which would be the background geometry) 
   for some turbulent region where most of the interesting dynamics are happening 
   (as we can not solve for an infinitely large grid due to restrictions imposed 
   on us by space and time complexity and computation power). 

   The field of numerical relativity is growing rampantly with the vast literature 
   on algorithms, numerical methods and theoretical formulations (from basic 3+1 
   decomposition formulation to more sophisticated ones) with the development of 
   robust and involved frameworks that provide a complete programming ecosystem 
   and have proved to be essential tools for any numerical relativity researcher.
   
   Reflecting the other end of the research community stands the major section 
   of theoretical physicists which counts the majority as a novice in programming. 
   The head challenge is the fact that the usage of these frameworks demands 
   heavy use of high-level programming languages like C, C++, Python, etc. 
   Though Python provides a vast room for abstractions, no library existed that  
   focused towards numerical relativity, eventually, the need of python library  
   dedicated to general relativity became the propellant for the team and thus 
   they made dauntless efforts to cover the space for this Python library. 
   Since then, EinsteinPy has seen a lot of contributions from people all over 
   the world and many ”good to go” functionalities are already provided in 
   this and previous versions.
   
   EinsteinPy is made to provide a set of tools which can make numerical 
   computations for solving Einstein’s field equations an easy job for anyone 
   who does not want to deal with intricacies of the subject along with few, 
   very basic but powerful functionalities that could be used by anyone who 
   wants to learn the subject of general relativity with every minute detail.
   
   The library is as an attempt to provide programming and numerical environment 
   for numerous numerical relativity problems like geodesics plotter, gravitational 
   lensing and ray tracing, solving and simulating relativistic hydrodynamical  
   equations, plotting of black hole event horizons, solving Einstein’s 
   field equations and simulating various dynamical systems like binary merger etc.
   
   EinsteinPy provides an open-source and open-development core package and 
   aims to build an ecosystem of affiliated packages that support numeric 
   relativity functionality in the Python programming language. EinsteinPy 
   core package is now a feature-rich library of sufficiently general tools 
   and classes that support the development of more specialized code.
   
   In the further coming section, we discuss with some of the features of the 
   current as well previous version and the plans which are yet to be implemented 
   in the upcoming versions. Also, we describe a few code snippets to explain 
   the usage of the library on which more details can be found on the 
   project website. We start this paper by describing the way EinsteinPy functions 
   followed by the main software efforts developed by the EinsteinPy itself 
   with proper detailed documentation of the current version release. We end 
   with a short vision for the future of EinsteinPy to its contribution in 
   the field of computational methodology in gravitational and 
   relativistic Astrophysics.

\section{Data Types}

  The EinsteinPy package provides some core data types for calculating, 
  operating, visualizing geodesics and black holes. There are some data 
  types created for symbolic manipulations and calculations of relativistic 
  tensors. The most important data types for EinsteinPy are \texttt{Body}, 
  \texttt{Geodesic} and \texttt{BaseRelativityTensor} objects that support 
  defining and creating new bodies, storing and calculating geodesic for 
  bodies, and handling respectively. The main purpose of these core classes 
  is to standardize the way calculations and visualizations are done for 
  various kinds of heavy cosmic bodies and black holes. The classes always 
  maintain the consistency in units of measurement wherever applicable, and 
  create a consistent and clean API for any non-programmer to understand. 
  The classes are designed such that it resembles the flow of physics and 
  are as intuitional as possible. These core classes also include a rigid 
  architecture for data manipulation and visualizations wherever applicable. 
  Following sections provide deep insight into the above mentioned three 
  core data types of EinsteinPy:
  
  \subsection{Body}
  The Body data type helps define an attractor i.e. the central black-hole 
  or the heavy mass object and later the objects under its influence. 
  The bodies can have their characteristics defined then and there thus 
  helping the user create the system easily. 
  
  \subsection{Geodesic}
  A geodesic is the path that an-accelerating particle would follow. 
  In the plane, the geodesics are straight lines. On the sphere, the 
  geodesics are great circles. The Geodesic data type helps define the 
  geodesic of a body in the presence of an attractor or more aptly, 
  according to the system user created. These \texttt{Geodesic} objects 
  can be defined for any configuration and metric. 
  
  \subsection{BaseRelativityTensor}
  This is a base class from which other tensors in \textit{symbolic} 
  module are derived. The user can use this class to create his/her 
  own tensor while using the various in-built functions provided by this 
  class. The class also maintains the indicial configuration of 
  the tensor(contravariant and covariant indices), calculates the 
  free variables used in describing the tensor other than those which 
  depicts the axis of space-time itself (e.g. $t, x, y, z$). The class 
  has a function to convert the symbolic tensorial quantities into 
  functions where actual numerical values can be substituted.

\section{Unit Handling}

All user-facing functionality provided by EinsteinPy make use of Astropy \citet{astropy:2013} units. All functions and objects must have their input 
constrained to the appropriate type of unit (e.g., length, mass or energy). 
Inputs can then be provided with any units that match the required type 
(e.g., millimeters and inches are both valid units for length) and conversions 
occur automatically without user intervention. The einsteinpy.constant 
subpackage contains a few standard constants.

\section{Development Model}
The EinsteinPy Project follows an open development model, which is widely used in the scientific Python community and other computational astrophysics packages. The project is hosted on GitHub as a public repository with restricted write access to some key people. However, anyone is welcomed and is encouraged to make code contributions and suggest changes using pull requests. Since the repository containing the codebase is licensed under an open and permissive MIT licence,  the project allows commercial use, modification, sub licensing, distribution and private use as long as the copyright notice and license are redistributed. The project uses git for distributed version control and relies on CircleCI, Codefactor, GitHub Actions etc. for running the unit tests and other CI checks for every pull request.
The code quality is maintained at the highest standards with a strong control on the cyclomatic complexity, code linting, imports ordering and a proper docstring wherever necessary. Every contribution is reviewed to further see if it fulfils the following requirements:

\begin{itemize}
\item \texttt The code must strictly adhere to the PEP8 Standards, on the top of it, the code must be lined with black to save the reviewer’s time identifying the changes made and reduce the diff generated because of the patch.
\item \texttt The imports in the code must be sorted using tools such as isort. 
\item \texttt The documentation follows numpy like docstrings, and a proper reStructured Text syntax is followed for writing the documentation.
\item \texttt The changelog is populated at release-time by the lead developer and the releasing developer
\end{itemize}
EinsteinPy further follows a trunk-based development model, which ensures a new branch for every minor release and new patch release made using the minor release branch. All the latest changes are kept on the master branch on GitHub and the relevant issues are closed as soon as the pull request solving it merges onto the master. A release having the changes is made at a later date. 
EinsteinPy only supports the latest bleeding-edge Python, Numpy\citep{numpy} and Astropy \citep{2019arXiv190710121V}{astropy:2018} versions.  All the stable Python versions after Python 3.5 are supported. All the code snippets written follow the guidelines of numba \citep{Lam2015NumbaAL} to accelerate the Python code and make the simulations faster. 

\section{The EinsteinPy Core Package} 
This section explains in detail, the modules available within EinsteinPy core package, their API and the purpose they serve.
\subsection{Metric}
This module captures all the geometric and causal structure of specific space-times, and can calculate trajectories in a given space-time. 
The module derives its name from Metric Tensor, a tensorial quantity used to describe differential lengths, especially in curved space-time. 
However, this module serves more than its usual definition and ultimately leads us to obtain geodesics of particles in a space-time with high curvature, where notions of Newtonian physics fail.
\subsubsection{Schwarzschild Metric} 
Karl Schwarzschild pioneered the analysis of the relation between the size of black hole and its mass and this work paved the way for the first exact solution of Einstein's Field Equation under the limited case of single spherical non-rotating, non-charged body. 
The metric equation \citep{Schwarzschild:1916uq} corresponding to a space-time centred around a body in the above described conditions is given by,

\begin{align}
& ds^2 = -c^2 d\tau ^2 = - (1 - \frac{r_s}{r})c^2 dt^2  + \frac{1}{(1 - \frac{r_s}{r})}dr^2 + r^2 d\theta ^2 + r^2 sin^2 \theta d\phi ^2
\end{align}

Schwarzschild Radius is a limit below which the degree of space-time curvature causes the particle to undergo irreversible gravitational collapse wherein the particle would need a speed greater than speed of light to escape singularity.

As $r \to \infty$, the metric defined above starts to approximate flat Minkowski space-time, thus showing that Schwarzschild geometry is asymptotically flat. 

If supplied with initial position and velocity of a particle with negligible mass(in comparison to the central body), we can obtain the geodesics of the particle for a given range of proper time $\tau$, as it boils down to a set of eight differential equations numerically solvable by a class of Runge-Kutta methods.

\begin{figure}[ht]
	\centering
	\includegraphics[scale=0.7]{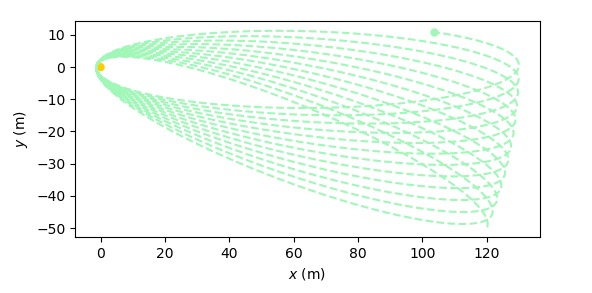}
	\caption{Plot obtained from the above code showing the phenomenon of perihelion advance}
	\label{fig:schwarzschild}
\end{figure}

We observe the phenomenon of perihelion advance of a particle around a heavy body due to spacetime curavture. The orbit would have been purely eliptical if plotted in accordance with Newtonian Mechanics, however, the curvature of space-time around the heavy mass shifts the orbit with each revolution. In practicality, this phenomenon is not so pronounced as no particle approaches a heavy mass this closely. For instance, Mercury advances its perihelion at a rate of $42.98\pm 0.04$ arcsec/century.

\subsubsection{Kerr Metric}
Kerr metric \citep{Kerr:1963ud} is a further generalization of Schwarzschild metric and a specific case of Kerr-Newman geometry. A space-time is defined as a "Kerr" space-time as and when the massive central body possesses an angular momentum. The corresponding metric equation which defines the space-time is,

\begin{gather}
g_{\mu \nu} = \begin{bmatrix} - (1 - \frac{r_s r}{\Sigma})c^2 & 0 & 0 & -\frac{r_s r a sin^2 \theta}{\Sigma}c \\ 0 & \frac{\Sigma}{\Delta} & 0 & 0 \\ 0 & 0 & \Sigma & 0 \\ -\frac{r_s r a sin^2 \theta}{\Sigma}c & 0 & 0 &  (r^2 + a^2 + \frac{r_s r a^2}{\Sigma}sin^2 \theta) sin^2 \theta \end{bmatrix} \\ 
\nonumber \\
ds^2 = g_{\mu \nu} dx^\mu dx^\nu
\end{gather}

where,
\begin{gather}
a = \frac{J}{Mc}, \nonumber \\
\Sigma = r^2 + a^2 cos^2, \nonumber \theta \\
\Delta  = r^2 - r_s r + a^2 \nonumber
\end{gather}

$J$ is the angular momentum of the body and all the other symbols hold their usual meanings as discussed in \ref{subsubsec:sch}. The coordinate system used to describe the metric is Boyer-Lindquist coordinate system.

APIs for \textit{Schwarzschild}, \textit{Kerr} and \textit{KerrNewman} are consistent and code structure remains same for the sake for clarity and intuitiveness. For example,

\begin{figure}[ht]
	\centering
	\includegraphics[scale=0.5]{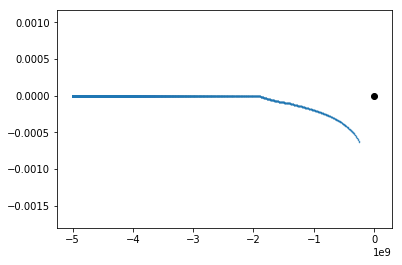} 
	\caption{Plot obtained from the above geodesics showing the phenomenon of frame dragging}
	\label{fig:kerr}
\end{figure}

\subsubsection{Kerr-Newman Metric}
Kerr-Newman metric \citep{Adamo:2014baa} presents the most general case within single-body vacuum solutions of Einstein Field Equation. It generalizes all the quantities used to describe a black hole, as affirmed by "no hair" theorem, i.e. Charge $Q$, Angular Momentum $J$ and Mass $M$. The metric equation is given by,

\begin{gather}
-ds^2 = c^2 d\tau^2 = -(\frac{dr^2}{\Delta}+d\theta^2)\rho^2 + (c dt - a sin^2 \theta d\phi)^2 \frac{\Delta}{\rho^2} - ((r^2 + a^2)d\phi - ac dt)^2 \frac{sin^2 \theta}{\rho^2}
\end{gather}

where,

\begin{gather}
a = \frac{J}{Mc}, \nonumber \\
\Sigma = r^2 + a^2 cos^2 \theta , \nonumber  \\
\Delta  = r^2 - r_s r + a^2 + r_Q ^ 2, \nonumber \\
r_Q ^ 2 = \frac{Q^2 G}{4\pi\epsilon_0 c^4} \nonumber
\end{gather}

All the other symbols convey their usual meanings, as discussed in the previous two sections Schwarzschild and Kerr Metric. For the sake of brevity, we would not go into the specifics of this class as the classes within \textit{Metric} module share very similar APIs as inferred in section for \texttt{Kerr}.

\subsection{Coordinates}
The coordinates subpackage provides support for representing and transforming coordinates. EinsteinPy provides support in Cartesian, Spherical and Boyer-Lindquist Coordinate systems. All of these are inter-convertible and can be used anytime. All the calculations within the module are majorly done in Spherical system. Numba \citep{Lam2015NumbaAL} is used to speed up the coordinate conversions for an extra performance boost.
 
EinsteinPy is the only actively maintained Python project which supports Boyer-Lindquist Coordinate system \citep{doi:10.1063/1.1705193}. Boyer and Lindquist wanted to modify the Kerr metric in a form closer to the usual form of the Schwarzschild metric. To do this, they defined a new coordinate system(t, r, $ \theta, \phi$).  In spherical coordinates where surfaces of constant r are spheres, but in  Boyer-Lindquist coordinates the corresponding surfaces are ellipsoids.

\subsection{Symbolic}
General Relativity required heavy mathematical formulation to derive and describe various tensors attributing to various features essential for the cogent description of an arbitrary space-time. We would describe the various quantities currently supported by this module while maintaining brevity. Each quantity has its class, even scalars, which are treated as 0\textsuperscript{th} order tensors.

All the mathematical quantities build upon \textit{BaseRelativityTensor} as derived classes, which is discussed in detail in section describing \textit{BaseRelativityTensor}.

\subsubsection{MetricTensor}
Metric Tensor describes the differential length elements required to measure distance in a curved(also applies to flat) space-time. It is a second order tensor and is fundamental to describe any space-time geometry. The \textit{MetricTensor} class, being inherited from  \textit{BaseRelativityTensor} , inherits all its functions and also have some functions and constraints of its own. For example, metric tensor is bound to be a second order tensor.


Also, the library maintains an ever expanding predefined metrics for direct use in research.





\subsubsection{ChristoffelSymbols}\label{subsubsec:chl}

Although Christoffel Symbols belong to class of pseudo-tensors, it is inherited from \textit{BaseRelativityTensor} to maintain homogeneity within the module. Being inherited from \textit{BaseRelativityTensor}, it already supports the functions of its parent class and imposes some restrictions of its own, such as order of the tensor should be 3.



\subsubsection{RiemannCurvatureTensor}\label{subsubsec:riemann}

This quantity is the most extensive measure of curvature of a space-time. It is inherited from \textit{BaseRelativityTensor} and supports all the functions of its parent class. 



\subsubsection{RicciTensor}\label{subsubsec:riccitensor}

This tensor is used in solving Einstein's  equation and is basically contraction of Riemann Curvature Tensor.  It is inherited from \textit{BaseRelativityTensor} and supports all the functions of its parent class. 



\subsubsection{RicciScalar}\label{subsubsec:ricciscalar}

It is a scalar quantity obtained by contracting a given Ricci Tensor. It is inherited from \textit{BaseRelativityTensor} and supports all the functions of its parent class. Being a scalar quantity, it does not support \texttt{change\_config} unlike other tensors as discussed above.



Here is a sample code snippet, using the above discussed classes : 

\begin{lstlisting}[language=Python, caption=Importing Anti De-Sitter metric]
>>> from einsteinpy.symbolic.predefined import AntiDeSitter
>>> from einsteinpy.symbolic import RicciTensor, RicciScalar
>>> import sympy
\end{lstlisting}

\begin{lstlisting}[language=Python, caption=Obtaining Ricci Tensor from the metric]
>>> RT = RicciTensor.from_metric(AntiDeSitter())
>>> RT.tensor()
\end{lstlisting}

\begin{center}
$\displaystyle \left[\begin{matrix}3 & 0 & 0 & 0\\0 & - 3 \cos^{2}{\left(t \right)} & 0 & 0\\0 & 0 & \left(\sin^{2}{\left(t \right)} - 1\right) \sinh^{2}{\left(\chi \right)} - 2 \cos^{2}{\left(t \right)} \sinh^{2}{\left(\chi \right)} & 0\\0 & 0 & 0 & \left(\sin^{2}{\left(t \right)} - 1\right) \sin^{2}{\left(\theta \right)} \sinh^{2}{\left(\chi \right)} - 2 \sin^{2}{\left(\theta \right)} \cos^{2}{\left(t \right)} \sinh^{2}{\left(\chi \right)}\end{matrix}\right]$
\end{center}

\begin{lstlisting}[language=Python, caption=Deriving Ricci(Curvature) Scalar from Ricci Tensor]
>>> RS = RicciScalar.from_riccitensor(RT)
>>> sympy.simplify(RS.expr)  # simplifying the expression
\end{lstlisting}

\begin{center}
$\displaystyle -12$
\end{center}

In congruence to theoretical results, the code outputs a constant negetive scalar curvature for Anti De-Sitter spacetime. All the above discussed classes are called in this snippet as \textit{RicciTensor} internally calls \textit{RiemannCurvatureTensor}, which in turn calls \textit{ChristoffelSymbols}.

\subsubsection{Other Defined Quantities}\label{subsubsec:other}

As the library is in heavy development, new quantities are being added as and when required. Some other quantities currently supported are listed below. The quantities are all inherited from \textit{BaseRelativityTensor}, and a similar API (same as that of discussed above) has been implemented to maintain compatiblity and homogeneity. The class names denote usual quantities relevant to General Relativity.

\begin{itemize}
\item \textit{EinsteinTensor}
\item \textit{StressEnergyMomentumTensor}
\item \textit{WeylTensor}
\item \textit{SchoutenTensor}
\end{itemize}

\subsection{Hypersurface}
This module provides basic computational functions which are essential for modelling the space-like hypersurface for any space-time geometry. Currently, the module has an implementation for Schwarzschild geometry with conversion functions from Schwarzschild coordinates to 3-D spherical coordinates and some plotting utilities. The whole module is expected to extend for providing a super-class for implementations related to gravitational lensing simulations and numerical calculations of null geodesics on hypersurfaces and deviation angles for systems which are relevant to relativistic astrophysical problems. A small code snippet for plotting the hypersurface is as follows:
\begin{lstlisting}[language=Python, caption=Schwarzschild Embedding]
>>> surface_obj = SchwarzschildEmbedding(5.927e23 * u.kg)
>>> surface = HypersurfacePlotter(embedding=surface_obj, plot_type='surface')
>>> surface.plot()
>>> surface.show()
\end{lstlisting}
\begin{figure}[ht]
	\centering
	\includegraphics[scale=0.5]{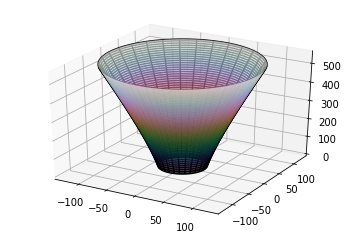} 
	\caption{Schwarzschild embedding}
	\label{fig:hypersurface}
\end{figure}

\subsection{Plotting}
The plotting ecosystem in EinsteinPy has two back-ends for two type of users. A static, matplotlib based plotter for publication-quality plots and an interactive plotly based plotter for analysis and interpretation. \\
The plotting module is specifically designed keeping in mind the variety of IDEs and tools people working with Python use, for example, python scripts, jupyter notebooks, spyder, pycharm etc. As plotly plots need JavasScript to render plots, it becomes difficult for the first time user to decide which backend to use. \texttt{StaticGeodesicPlotter} and \texttt{InteractiveGeodesicPlotter} undergo an internal switching depending on the environment variables and the appropriate environment aware plot is shown to the user for better user experience and interface.\\
The plotting module throughout the applications follow a very consistent API. \begin{lstlisting}[language=Python, caption=Plotting Module API]
>>> data = Plotter()
>>> data.plot()
>>> data.show()
\end{lstlisting}

\subsection{Rays}
The rays module of EinsteinPy tries to focus on how light behaves around heavy mass object.
The major functionality this module provides is the utility functions for carrying out shadow related computations for different black hole spacetimes. The module has natural data types to efficiently store impact parameters and use them in further computations related to wavelength shifts, turning point etc. The specific assumptions about the accretion process and the emission mechanisms are in accordance with that given in \citep{Bambi_2013}.
The module also provide functionalities related to intensities of blue shifted and red shifted rays above and below the critical impact parameters. For mathematical formalism of the above computations refer to the \citep{Bambi_2013} and \citep{Shaikh_2018}.

This module is under heave developement since past February 2020, and we expect significant contributions to this module thanks to Google Summer of Code 2020.

\section{Infrastructure} \label{sec:infra}
\subsection{Testing} \label{subsec:testing}
The EinsteinPy Project maintains a very high bar in accepting changes and introducing new features. Every patch of code that is accepted in the codebase through any pull request has to be well tested using some unit tests unless given an exception after discussion with both the lead developers. Currently, the coverage of unit tests, which means the total fraction of codebase run by at least one unit test is 94\% and we hope to improve it in the future. The 100\% code coverage for this software at this level is not possible because of the lack of accurate data to test the code against. We deliberately leave some parts of the codebase out of the testing loop because testing those modules is not possible with the current knowledge level and expertise. However, 94\% code coverage is good enough when we consider other packages in the scientific Python community who face a similar problem.  

The test suite can be run manually as well as it runs in an automated fashion on every commit as well as pull request to maintain the code standards, and make it easier for contributors to understand the implications and impact of their changes. The EinsteinPy Project uses many continuous-integration services which provide services to create automated pipelines for code inspection, unit testing, coverage publishing etc. All the contributions trigger these free services (Codecov, CircleCI, Appveyor, GitHub Actions, Codefactor, Codeclimate, Hound, WIP) which are integrated with GitHub repository. All the pull requests are checked for code linting by a pep8speaks bot, which suggests the lines of code which violate the PEP8 standard to the contributor in a human-readable way. All these services run on three types of operating systems (Windows, macOS and Linux). The unit tests run on various versions of Python. They test the documentation builds, test coverage, check whether the pull request author has indicated that it is a work in progress (WIP). They also run several code linting checks which further triggers the unit test build.

\subsection{ Documentation and gallery} \label{subsec:doc}
The EinsteinPy Project strives to provide high quality, up-to-date and accurate documentation of the software. The documentation is highly approachable with an easy and intuitive user interface. The documentation is self-updating with every change in the codebase and is rendered automatically from the docstrings written for each function and classes. Along with the main API, new example codes for different functionalities are added to the repository frequently in the form of jupyter notebooks. Many sections like User Guide and Developer Guide has been written for easy on-boarding of new users and contributors.
	
\section{Community} \label{sec:COMMUNITY}
EinsteinPy is being developed as an open-source community inside the wide and diverse general scientific python community. Our code is freely available, and we have seen a lot of people contributing and joining us since over the past year. We touched the landmark of 50 contributors recently. EinsteinPy is dedicated to problems arising in General Relativity and gravitational physics as well as encouraging open source development. EinsteinPy has benefited greatly from summer of code schemes. In 2019 EinsteinPy participated in the ESA Summer of Code In Space (SOCIS). This program is inspired by the Google Summer Of Code (GSoC) and aims to raise the awareness of open source projects related to space, promote the European Space Agency. The symbolic module in EinsteinPy project was developed during this program.

In 2020, EinsteinPy is selected as a sub-organization in Google Summer of Code under OpenAstronomy Umbrella. 

\section{Future} \label{sec:FUTURE}
We intend to add more functionalities to our project, some of which are, 
\begin{itemize}
\item Support for null-geodesics in different geometries, . 
\item Relativistic hydrodynamics
\item Adaptive Mesh Refinement 
\item Providing numerical solutions to Einstein’s equations for arbitrarily complex matter distribution.
\end{itemize}

We have an open community where we welcome suggestions from people to add functionalities/features and everyone can report bugs and issues. We wish to be as helpful as possible to people using EinsteinPy in their research.

\section{Conclusions}
The paper outlines the package EinsteinPy, a community and open effort for solving problems in general relativity. We discussed all the modules of the library:
\begin{itemize}
\item Metric Module
\item Coordinates Module
\item Plotting Module
\item Symbolic Module
\item Rays Module
\item Hypersurface Module
\end{itemize}

We discuss the basic development practices that EinsteinPy follows, and the future goals.

\begin{acknowledgements}
      We thank all the people that have made EinsteinPy what it is today.  
The generous sponsorship by European Space Agency during Summer of Code in Space really helped us with the betterment of code. We would like to thank the awesome poliastro \citep{juan_luis_cano_rodriguez_2019_3588160} for always motivating us and for being the first sponsor for the project. We would also like to thank our host university, Indian Institute of Technology, Mandi who provided us with significant support and have our deepest thanks.
We would especially like to thank Prof.Abhay Ashtekar for his motivational email and  Prof. Timothy A. Gonsalves for his constant motivation and support. We would also like to thank Prof. Arnav Bhavsar for his constant support.
Lastly, we thank all our contributors who contributed selflessly for building this amazing software called EinsteinPy.
      
\end{acknowledgements}

%
%

\bibliographystyle{aa}
\bibliography{bibliography.bib}

\end{document}